\documentclass[a4paper]{article}

\usepackage{19lomcon}        
\usepackage{cite}             
\usepackage{epsfig}           
\usepackage{epstopdf}

\begin{document}


\title{HISTORY OF NEUTRINO MAGNETIC MOMENT}

\author{Jihn E. Kim \email{jihnekim@gmail.com} 
}

\affiliation{Department of Physics, Kyung Hee University, 02447
Seoul, Republic of Korea \\~\\  }


\date{}
\maketitle


\begin{abstract}
In this historic Lomonosov conference on the occasion of 150 year anniversary of the Mendeleev's periodic  table, I present the history of neutrino magnetic moment. It was first thought  by Wolfgang Pauli and its magnitude was calculated  during the gauge theory era. 

\end{abstract}

\section{Introduction}\label{sec:Introduction}

In neutrino physics, the magnetic moment of neutrino is customarily expressed as $f$ times the Bohr magneton,
\begin{equation}
\mu = f\cdot 1.93\times 10^{-11\,}e\,{\rm cm}=f\,\mu_{_B} 
\end{equation}
where
\begin{equation}
\mu_{_B}=\frac{e\hbar}{2m_e}=\frac{\sqrt{4\pi\alpha_{\rm em}}\,\hbar}{2m_e}=0.2963\,{\rm MeV}^{-1}=\frac{1}{3.375\,{\rm MeV}} 
=0.5846\times 10^{-11\,}{\rm cm}.
\end{equation}

The neutrino magnetic moment was first commented by Pauli in his famous letter of proposing the neutrino to the audience of the T$\ddot{\rm u}$bingen meeting of the German Physical Society, with the date on the letter December 4, 1930, 
``Dear Radioactive Ladies and Gentlemen, .........~In more detail, how because of the ``wrong''  statistics of the N and $^6$Li nuclei and the continuous beta-spectrum, I have hit upon a desperate remedy to save the
``exchange theorem'' of statistics and the law of conservation of energy. Namely, the possibility that there could exist in the nuclei electrically neutral particles, that I
wish to call neutrons, which have spin 1/2 and obey the exclusion principle and which further differ from light quanta in that they do not travel with the velocity of light.......~Now the question
that has to be dealt with is which forces act on neutrons? The most likely model for
the neutron seems to me, because of wave mechanical reasons (the details are
known by the bearer of these lines), that neutron at rest is a magnetic dipole of a
certain moment mu. The experiment seems to require that the effect of  the
ionization of such a neutron  cannot be larger than that of a gamma-ray and then
mu should not be larger than $e\times 10^{-3}$\,cm ......'' We know that his neutron is our neutrino now since the time  Enrico Fermi  named it as {\it neutrino} (a little neutral one) after the discovery of neutron in 1932. 

In 1930 when neutron was not known, there were problems in $\beta$-decay. Nitrogen of atomic number 7 and mass number 14 has spin-1 and the less abundant Lithium isotope of atomic number 3 and mass number 6 has also spin-1. The deuteron with atomic number 1 and mass tumber 2 has spin-1 also.  At that time electron was known. So, to make $^6$Li nucleus, one needs 6 protons and three electrons to make a $Q_{\rm em}=3e$ nucleus. With the odd number of fermions, in total 9, one cannot make a spin-1 nucleus. These were the problem observed in $\beta$-decays that was mentioned in Pauli's letter. Now we understand that there is an additional spin-$\frac12$ particle inside nuclei, neutron, and there is no problem. But in 1930, it was a big problem. Pauli thought that  in $\beta$-decays his neutron (our neutrino, and we will call his neutron as neutrino from now on) goes out of the mother nucleus but he knew only the electromagnetic interaction. That was the reason that even if neutrino is neutral it must have some electromagnetic property so that the  $\beta$-decays are possible. This was the first introduction of the magnetic moment of neutrino. Therefore, his estimate on the neutrino magnetic moment is absurdly large.

After Pauli, neutrino magnetic moment was studied experimentally in  1957, restricting  $f<10^{-9}$ \cite{Reines57}. This experimental bound has led to a phenomenological study of neutrino magnetic moment in 1963 \cite{Lee63}. Ten years later in the gauge theory era, the forward peak cross section of the electron neutrino on nucleon,  due to the $1/q^2$ propagator effects, was calculated to apply to the Solar neutrino problem \cite{Clark73}. Since late 1973, I published on neutrino magnetic moment every two years \cite{Kim74, Kim76, Kim78} until the strong CP problem became my favorate topic \cite{Kim79}. Ten years ago, I generalized this idea for estimating the WIMP production rate from the WIMP magnetic moment $f_M$ \cite{Kim10,Barger11}, to apply to the CDMS II experimental result of that present \cite{CDMSII09},
\begin{eqnarray}
\frac{d\sigma}{dE_{\rm rec}}=\frac{2\pi\alpha_{\rm em}^2f_N^2}{Mm_N^2|\vec{p}|^2}\Big[Z^2\left\{\frac{\Lambda_-(s,m_N^2,M^2)}{2ME_{\rm rec}} +(2m_N^2+M^2-s)\right\} +2ZF_2 \nonumber\\
\cdot(4m_N^2 -ME_{\rm rec})
 +F_2^2\left\{\frac{\Lambda_+(s,m_N^2,M^2)}{M^2} -\frac{2sE_{\rm rec}}{M}+E_{\rm rec}^2\right\}\Big]
\end{eqnarray}
where $Z, m_N,M,s$, and $E_{\rm rec}$ are, respectively, the atomic number,  the WIMP mass, the nucleus mass, the total energy and the recoil energy of the nucleus, and $f_N$ is the fraction of the magnetic moment (new Bohr magneton with $m_e\to m_N$) of mass $m_N$ particle of one unit charge.

\section{Neutrino Magnetic Moment}

For a Dirac neutrino, the chiral properties of neutrino magnetic
moment and neutrino mass are the same. Therefore,  the magnetic moment of a Dirac neutrino is in general only of O($eg^2$) times its mass.

Dirac neutrino's L-chirality and R-chirality are
combined and Lepton number can be defined. On the other hand,
for a Majorana neutrino, we use two L-chiralities or two R-chiralities in the mass term, and 
the Lepton number must be
broken if it is properly defined initially. But for a massless Majorana neutrino, it is identical to a Weyl neutrino which also has a half degrees of those of a Dirac neutrino.

In 1957, McLennan commented in \cite{McLennan57} that the two component theory of free neutrino in the parity non-conserving world is identical to the theory of Majorana neutrino. This could have led to a ``V--A'' theory of neutrinos by extending to all interactions of the neutrinos. In fact, for a massless neutrino, we now know that a two-component neutrino of the Weyl-type and of the Majorana-type are the same. But if non-zero mass is allowed, a Weyl neutrino must invite a two component neutrino such that it gets a Dirac mass. McLennan used the $\gamma_5$ invariance, so did Marshak and Sudarshan. The difference of these two groups was that the former restricted on the neutrino property only but the latter applied to all the weak interactions arriving at the ``V--A'' charged current interactions together with Feynman and Gell-Mann.

\section{Neutrino Magnetic Moment in the Gauge-theory Era}

The neutrino magnetic moment in gauge theory era is extensively reviewed  in Ref. \cite{Giunti15}, where the matrix elements of the QED interaction between neutrino states is defined as
\begin{equation}
{\cal H}^\nu=\overline{\nu}(x) \Lambda_\mu \nu(x) A^\mu(x)
\label{DefLambda}
\end{equation}
where in case of a Majorana neutrino $\overline{\nu}$ is written as $\nu^T C^{\dagger}$.

In the Standard Model (SM), the fifteen chiral fields do not allow Dirac magnetic moment since there is no $Q_{\rm em}=0$ singlet.
 One should add a  $Q_{\rm em}=0$ singlet which can be interpreted as a singlet R-handed field. Then, there can be a Dirac magnetic moment. Without a singlet, the neutrino in the SM can have an effective magnetic moment below the SM symmetry breaking scale but of the form of the Majorana type.

\subsection{Dirac neutrino}\label{subsec:Dirac}
In the case of Dirac neutrinos, $\Lambda_\mu$ in Eq. (\ref{DefLambda}) becomes
\begin{equation}
\Lambda_\mu^{fi}(q)= \left(\gamma_\mu -\frac{q_\mu {q\hskip -0.17cm \slash} }{q^2} \right)\left(f_Q^{fi}(q^2) +f_A^{fi}(q^2) q^2\gamma_5 \right)
-i\sigma_{\mu\nu}q^\nu\,f_M^{fi}(q^2)
\end{equation}
while the electric dipole moment term $\sigma_{\mu\nu}q^\nu\,f_E^{fi}(q^2)\gamma_5$  is forbidden by the condition of CP invariance. The above form factors are effectively real, $f_{\Omega}=(f_{\Omega})^\dagger$ for $\Omega=Q, M, E, A$.

\begin{figure}[t]
\centering
\includegraphics[height=4cm]{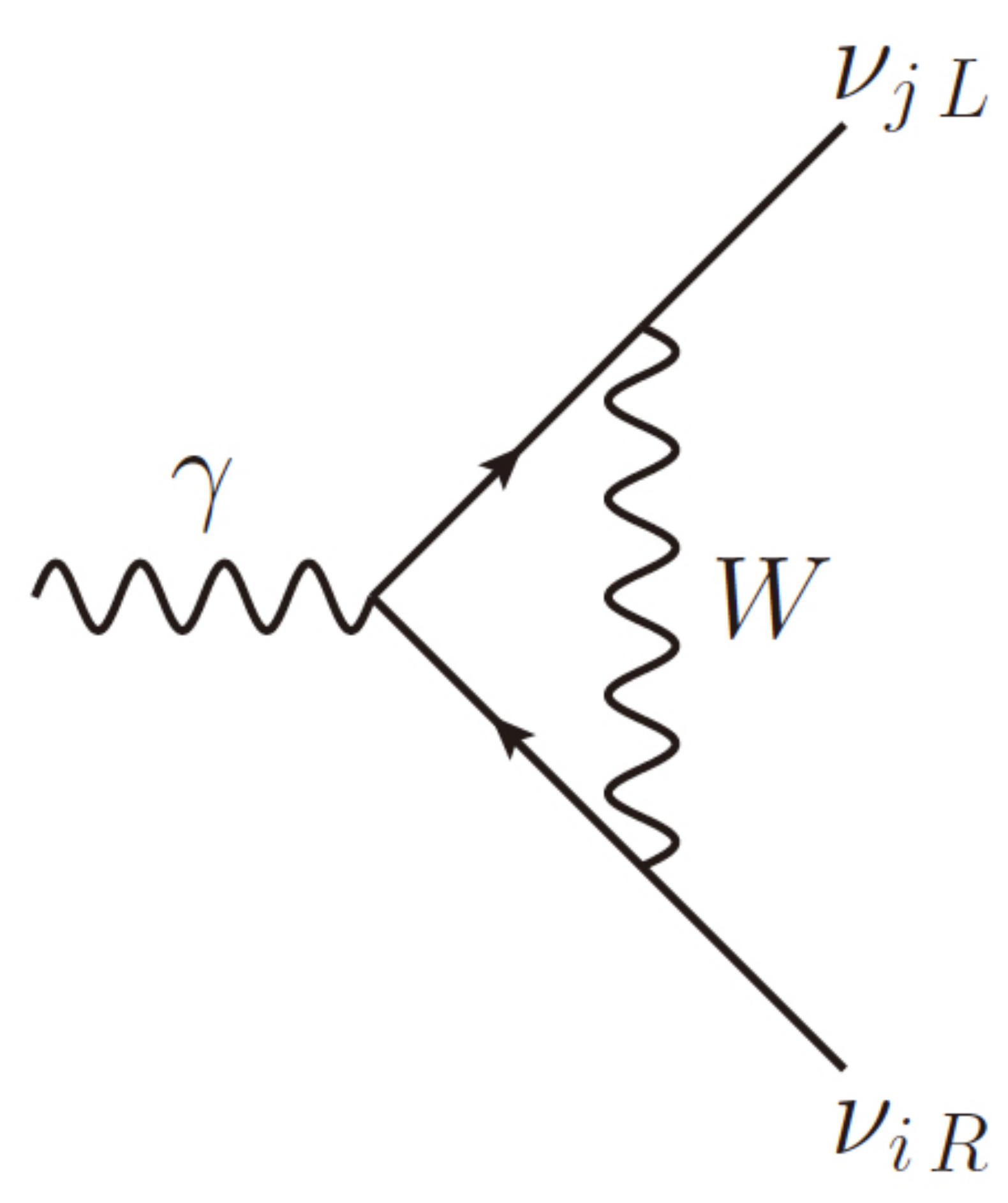}
\caption{Magnetic moment diagram for Dirac neutrinos.}
\label{fig:Dirac}
\end{figure}

The magnitude can be large, even of order 10$^{-4}$ \cite{Shrock74,Kim76,Sanda77,Kim78,Shrock82}.
Mostly, it is proportional to the neutrino mass,
multiplied by dimensionless number. It is pointed out that  there are more possibilities with charged Higgs \cite{Studenikin04}.  

\subsection{Majorana neutrino}\label{subsec:Majorana}
\begin{figure}[b]
\centering
\includegraphics[height=4cm]{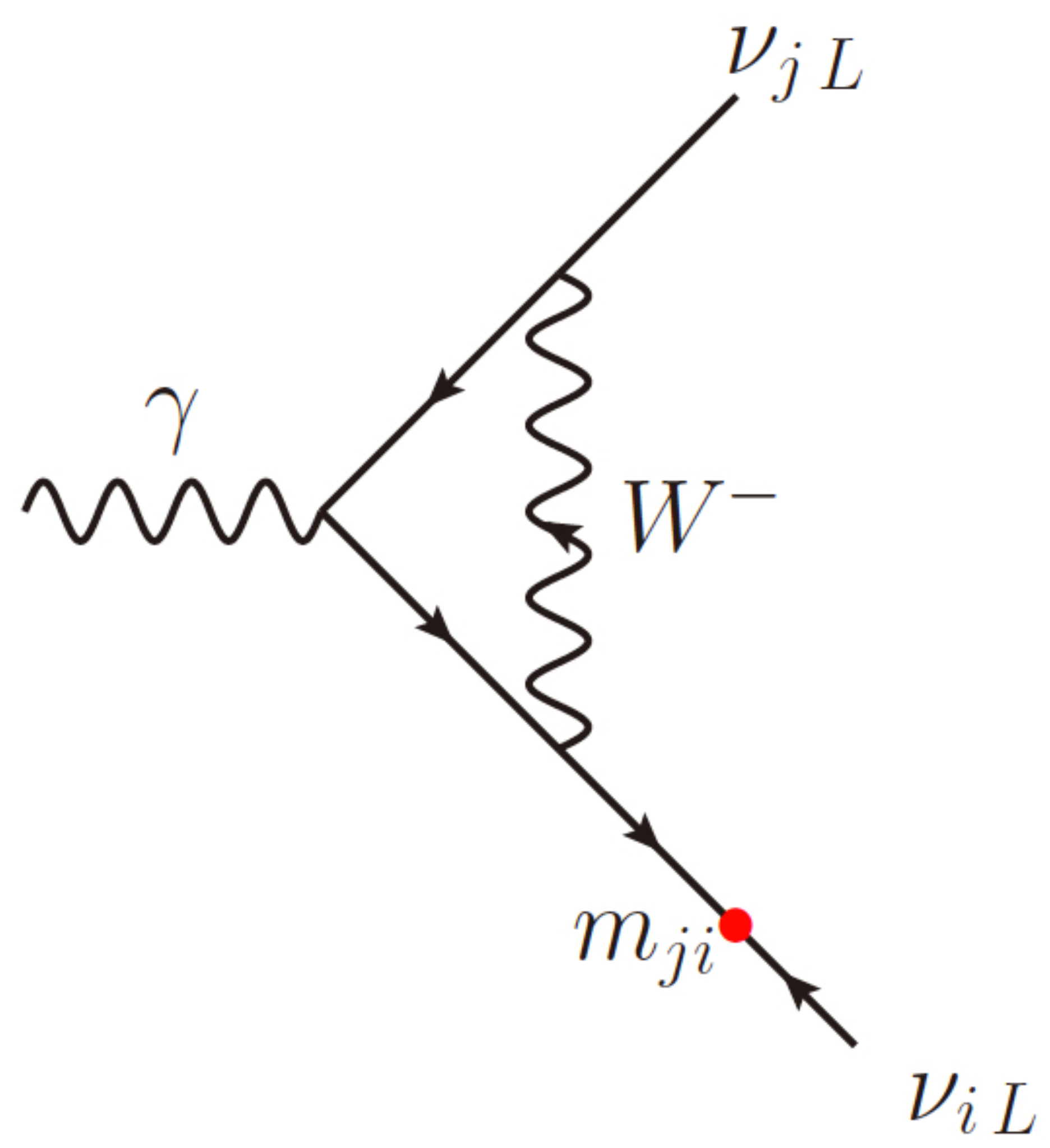}
\caption{Magnetic moment diagram for Majorana neutrinos.}
\label{fig:Majorana}
\end{figure}
In case of Majorana neutrinos,  $\Lambda_\mu$ in Eq. (\ref{DefLambda}) becomes
\begin{eqnarray}
\Lambda_\mu^{M}(q)= \left(\gamma_\mu -\frac{q_\mu{q\hskip -0.17cm \slash}}{q^2} \right)\left(f_Q^{M}(q^2) +f_A^{M}(q^2) q^2\gamma_5 \right)\nonumber\\
-i\sigma_{\mu\nu}q^\nu[f_M^{M}(q^2)+i f_E^{M}(q^2)\gamma_5]
\end{eqnarray}
where  $f^M_{A}=(f^M_{A})^T$  and  $f_{\Omega}=-(f_{\Omega})^T$ for $\Omega=Q, M, E$. In Fig. \ref{fig:Majorana}, the one-loop generation of the transition magnetic moment of neutrinos $\Lambda_\mu^{ij}(q)$ is shown. Here, the insertion of L-number violating Majorana neutrino mass $m_{ji}$ is shown. In the SM, Fujikawa and Shrock estimated the magnitude as
\begin{equation}
\mu_\nu=\frac{3eG_F\,m_\nu}{8\sqrt{2}\,\pi},
\end{equation}
which is about $f=3\times 10^{-21}$ for $10^{-2}\,$eV neutrino mass \cite{Fujikawa80}.

\subsection{Charge Radius and Magnetic Moment}\label{subsec:ChargeR}
The form factor coefficients can be written to lead
to an effective Lagrangian, by replacing $q_\nu\to i\partial^\nu$ followed by $A^\mu$, specifically,
\begin{equation}
-i\sigma_{\mu\nu} q^\nu A^\mu\to \sigma_{\mu\nu}\partial^\nu A^\mu=-\frac12 \sigma_{\mu\nu}F^{\mu\nu}
\end{equation}
which is gauge invariant. Therefore, the neutrino magnetic moment has a proper meaning even in the point particle limit. 

On the other hand, the charge radius is given by
\begin{equation}
\langle r^2\rangle\equiv 6\frac{df_Q(q^2)}{dq^2}\Big|_{q^2=0}
\end{equation}
The charge radius is gauge dependent because $A^\mu$ in Eq. (\ref{DefLambda}) couples to $\Lambda_\mu\sim \gamma_\mu$ and the resulting effective interaction is gauge dependent. Still the charge radius has a physical meaning if we consider it below the scale where we calculate the loop diagram as pion charge radius is
discussed below a few hundred MeV scale. In
our case it is below the $W^\pm$ mass scale. It will be better if the calculation does not depend on the gauge
choice. It has been discussed that specific calculations of the charge radius of neutrino  depended on the choice of gauge \cite{Bardeen72,LeeSY72,LeeShrock77}. So, the definition of charge radius and the follow-ups were worked out  by many people \cite{Zepeda85,Degrassi89,Bernabeu00}.

\subsection{Ultra-violet Completion}\label{subsec:Ultra}

In compactification models from some ultra-violet completed theory such as  the E$_8\times$E$_8'$ heterotic string, there
can be many U(1)’s. So, the neutrino magnetic moments can come from the kinetic mixing also. Usually the kinetic mixing parameter $\chi$ is of order $10^{-3}< \chi< 10^{-2}$ \cite{Dienes97}.

\begin{figure}[t]
\centering
\includegraphics[height=4.5cm]{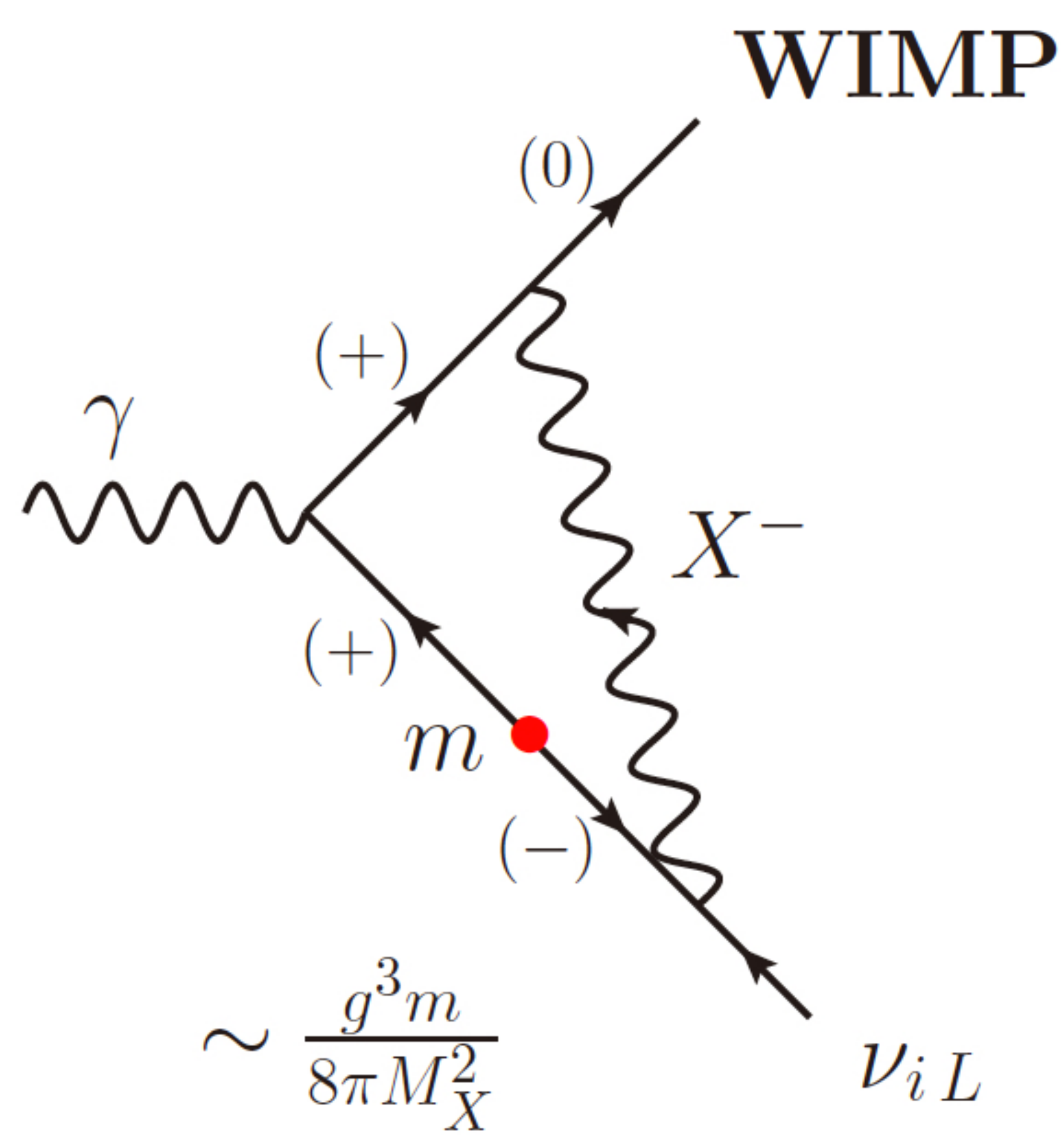}
\caption{A transition magnetic moment diagram to WIMP from a Dirac neutrino.}
\label{fig:WIMP}
\end{figure}

For the Majorana neutrinos, small magnetic moments are inevitable.
 But, for the Dirac neutrinos, however, there can be even a transition magnetic moment to WIMP  as shown in Fig. \ref{fig:WIMP}.

\section{Effects in stars}

The neutrino magnetic moment gives a helicity flipping interaction. Therefore, a L-handed neutrino becomes an R-handed neutrino through the magnetic moment interaction. The R-handed neutrinos are not participating the charged current weak interaction, which can deplete the L-handed electron neutrinos produced in the Sun. 

Clark and Pedigo pointed out in 1973 that it required for the magnetic moment of electron neutrino of $2.5\times 10^{-4}$  times Bohr magneton,  by calculating the forward
scattering amplitude, is needed to reduce the observable Solar neutrino flux to 50\% of the core-produced neutrinos \cite{Clark73}, which was  also pointed out ``possible'' in gauge theory models with the Dirac neutrino \cite{Kim76}.

In 1986, Okun, Voloshin, and Vysotsky applied this helicity flip of neutrinos by the magnetic moment interaction to the neutrino
propagation in the Sun for the study of Sunspot activity  \cite{Okun86}. Here, the attention is given to a neutrino's electromagnetic momentum
matrix and the possibility that some of its elements are of the
order of 10 to the $10^{-10}$ times Bohr magneton.  Also, they studied the flavor oscillations, i.e. transition magnetic moments, and the neutrino spin precession in a magnetic field in the presence of matter are examined. In their study, the interaction
between Solar neutrinos and the magnetic field inside the Sun's
convective zone could lead to 11-yr and semiannual flux
variations which are correlated with the Solar magnetic activity.

\section*{Acknowledgments}

This work is supported in part by the Korea National Research Foundation Grant  NRF-2018R1A2A3074631.


\end{document}